\documentclass[a4paper,11pt]{article}

\usepackage{pos}
\usepackage{multirow}
\usepackage{xcolor}

\title{Search for TeV Neutrinos from Seyfert Galaxies in the Southern Sky using Starting Track Events in IceCube
}

\ShortTitle{Southern Hemisphere Seyfert Galaxy Analysis using Starting Track Events with IceCube}

\author{The IceCube Collaboration \\{\normalsize \normalfont(a complete list of authors can be found at the end of the proceedings)}\\}

\emailAdd{shiqiyu@icecube.wisc.edu}

\abstract{

Supermassive black holes (SMBHs) power active galactic nuclei (AGN). The vicinity of the SMBH has long been proposed as the potential site of particle acceleration and neutrino production. Recently, IceCube reported evidence of neutrino emission from the Seyfert II galaxy NGC 1068. The absence of a matching flux of TeV gamma rays suggests that neutrinos are produced where gamma rays can efficiently get attenuated, for example, in the hot coronal environment near the SMBH at the core of the AGN. Here, we select the intrinsically brightest (in X-ray) Seyfert galaxies in the Southern Sky from the BAT AGN Spectroscopic Survey (BASS) and search for associated neutrinos using starting track events in IceCube. In addition to the standard power law flux assumption, we leverage a dedicated disc-corona model of neutrino production in such an environment to improve the discovery potential of the search. In this contribution, we report on the expected performance of our searches for neutrinos from these Seyfert galaxies.

\vspace{4mm}
{\bfseries Corresponding authors:}
Shiqi Yu$^{1*}$, Ali Kheirandish$^{2}$, Qinrui Liu$^{3}$, Hans Niederhausen$^{1}$ \\
{$^{1}$ \itshape Michigan State University}\\[4mm]
{$^{2}$ \itshape University of Nevada, Las Vegas}\\[4mm]
{$^{3}$ \itshape Queen's University}\\[4mm]
$^*$ Presenter

\ConferenceLogo{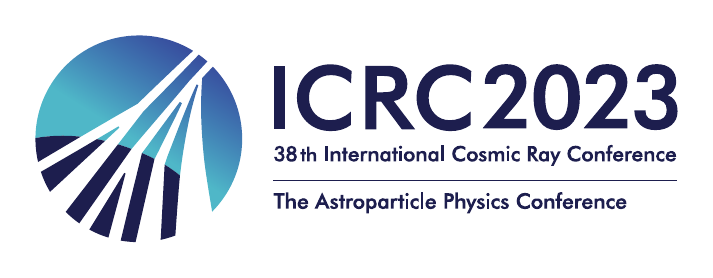}

\FullConference{The 38th International Cosmic Ray Conference (ICRC2023)\\ 26 July -- 3 August, 2023\\ Nagoya, Japan}
}

\begin{document}

\maketitle

\section{Introduction}\label{sec_introduction}

\noindent The IceCube Neutrino Observatory at the South Pole recently reported evidence of TeV neutrino emission from the nearby active galaxy, NGC 1068 \cite{IceCube:2022der}. This reinforces the idea that AGN could contribute to the diffuse flux of high-energy neutrinos observed by IceCube since 2013 \cite{IceCube:2013low}. Assuming the flux of neutrinos from NGC 1068 to follow a single power-law, the inferred muon neutrino flux at neutrino energy $E_\nu=1\,\mathrm{TeV}$ is $5\times 10^{-14} \, \rm GeV^{-1}\, cm^{-2}\, s^{-1}$ with spectral index of $3.2$ \cite{IceCube:2022der}. The corresponding neutrino luminosity in the $1.5\,\mathrm{TeV}$ to $15\,\mathrm{TeV}$ energy range is much higher than the reported GeV-scale $\gamma$-ray luminosity by the {\em Fermi} Large Area Telescope (LAT)~\citep{Fermi-LAT:2019yla,Ballet:2020hze} and the neutrino flux exceeds TeV-scale gamma-ray upper limits reported by MAGIC~\cite{MAGIC:2019fvw}.\\
The cosmic-ray (CR) interactions that produce neutrinos are expected to also generate a matching $\gamma$-ray signature. The absence of TeV gamma-rays from NGC 1068, therefore, suggests that the production site of high-energy neutrinos must be opaque to high energy $\gamma$-rays. Such conditions can be found in the matter and radiation-dense region in the closest vicinity of the central SMBH, commonly referred to as AGN corona, a component of the AGN that is known for its brightness in the X-ray band \cite{Inoue_2019,Inoue:2019yfs,Murase:2019vdl,Eichmann:2022lxh}. Coronae are generally assumed to be composed of a hot, highly magnetized, and turbulent plasma in the inner region of the accretion disk that is supported by accretion dynamics and magnetic dissipation ~\cite{Miller:1999ix}. The comptonization of soft photons from the accretion disc is thought to produce the observed X-rays. If protons are accelerated in this environment, for example, by turbulent magnetic fields, the Seyfert galaxies that are intrinsically bright in X-rays may also shine bright in neutrinos \cite{Kheirandish:2021wkm}. In this work, we refer to the model presented in \cite{Murase:2019vdl, Kheirandish:2021wkm} as the \textit{disc-corona model}.\\
Assuming parameters that could explain the neutrino flux from NGC 1068, and following \cite{Kheirandish:2021wkm}, we use this model to relate the intrinsic X-ray fluxes to neutrino production for the brightest Seyfert galaxies reported by BASS in the Southern Sky ($\delta<-5\,^\circ$). We then use down-going starting track events \cite{Silva:2023icrc} recorded by IceCube to search for the corresponding neutrino emission from these objects. In other words, in this work, we study sources that may be similar to NGC 1068 as far as the production of neutrinos is concerned. Experimental results from a related IceCube study, applying the same search strategy to Northern Seyfert galaxies while using through-going neutrino-induced muon tracks in the Northern Sky, are presented in a separate contribution \cite{IceCube:2023northseyfert}. 

\section{Neutrino Dataset and Source Selection}
\label{sec_selection}
\noindent Recent progress in the background rejection and reconstruction methods for starting events in IceCube, cascades \cite{Abbasi:2023bvn} and starting tracks \cite{Silva:2021pos:estes, Silva:2023icrc}, has significantly improved the detector's sensitivity to neutrino sources in the Southern Sky. Here, we analyze starting track events recorded by IceCube during a $10.3$ year period from May 2011 to January 2022. The event selection criteria and event reconstruction methods are discussed in \cite{Silva:2021pos:estes, Silva:2023icrc}. $10,350$ events pass all selection criteria, of which $2,091$ events are down-going, i.e., originate from the Southern Sky, and therefore contribute to this analysis. 90\% of these events have estimated neutrino energies ranging from $1.4\,\mathrm{TeV}$ to $29\,\mathrm{TeV}$. The median angular resolution in the Southern Sky ranges from $\sim 2\,\mathrm{deg}$ at a neutrino energy of $1\,\mathrm{TeV}$ to $\sim 0.4\,\mathrm{deg}$ at $1\,\mathrm{PeV}$. For a detailed discussion of this neutrino dataset and its performance, see \cite{Silva:2021pos:estes, Silva:2023icrc}.
\begin{figure}[tbp!]
\centering
\includegraphics[width=0.53\textwidth]{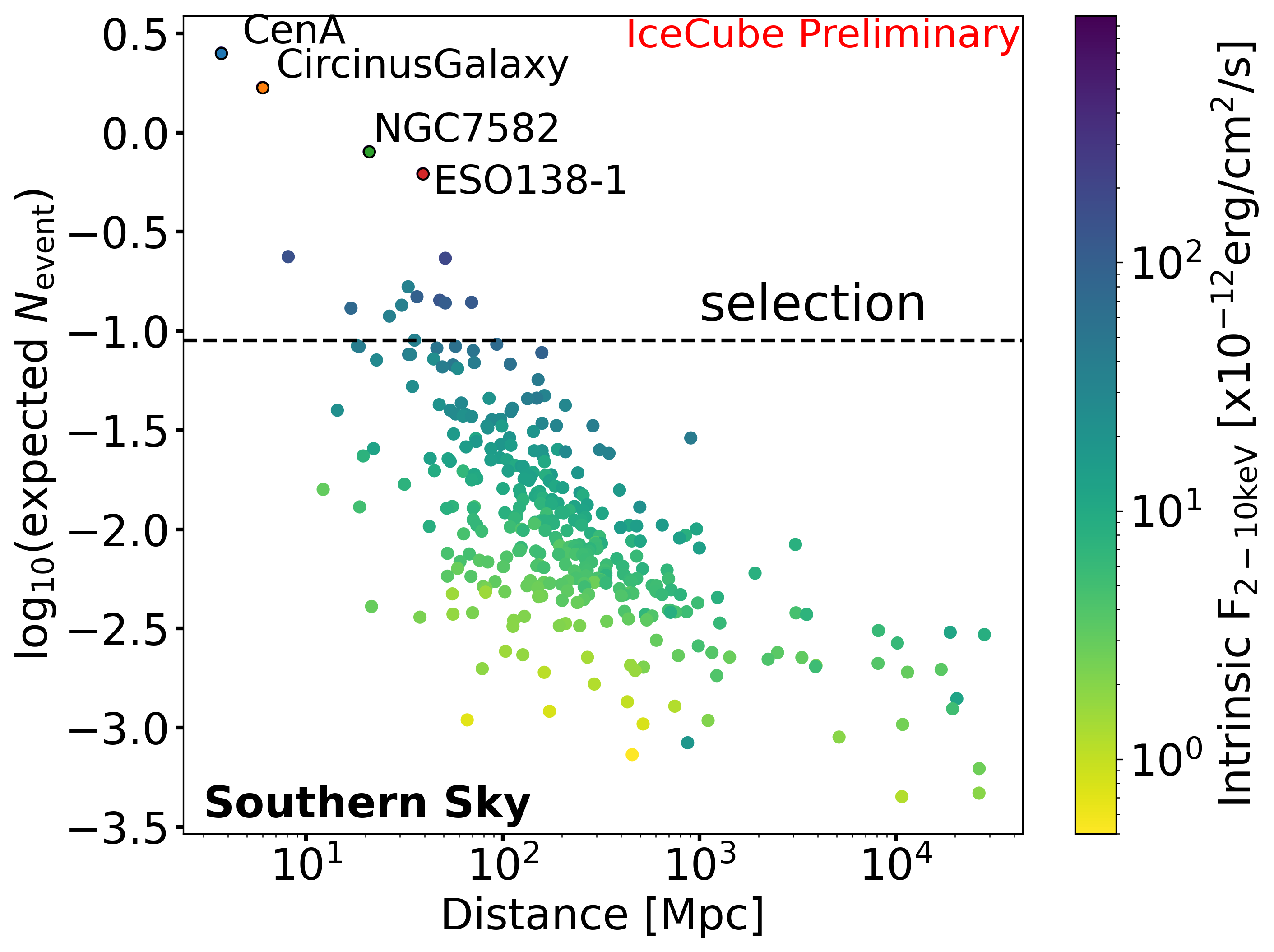}
\includegraphics[width=0.45\textwidth]{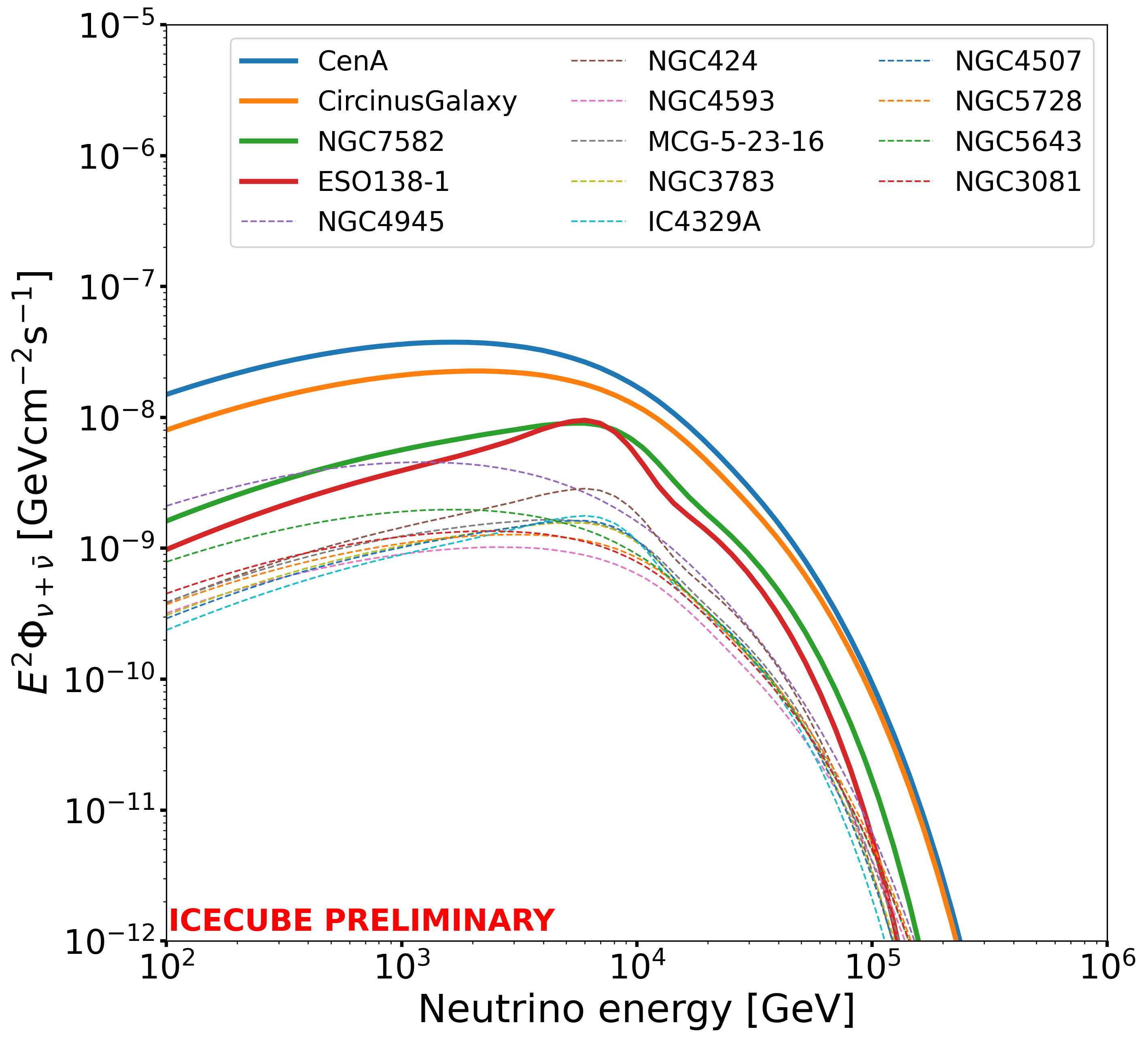}
\caption{Left: intrinsic flux at 2--10\,keV of Seyfert galaxies in Southern Hemisphere plotted against neutrino expectations by using disk-corona model flux vs.\ distance of the source from Earth, where selected sources above the dashed line. Right: disk-corona model predicted neutrino spectral flux of selected catalog sources with the top four predicted brightest sources highlighted as solid curves.}
\label{fig:selection}
\end{figure}\\

\begin{figure}[tbp!]
\centering
\includegraphics[width=0.8\textwidth]{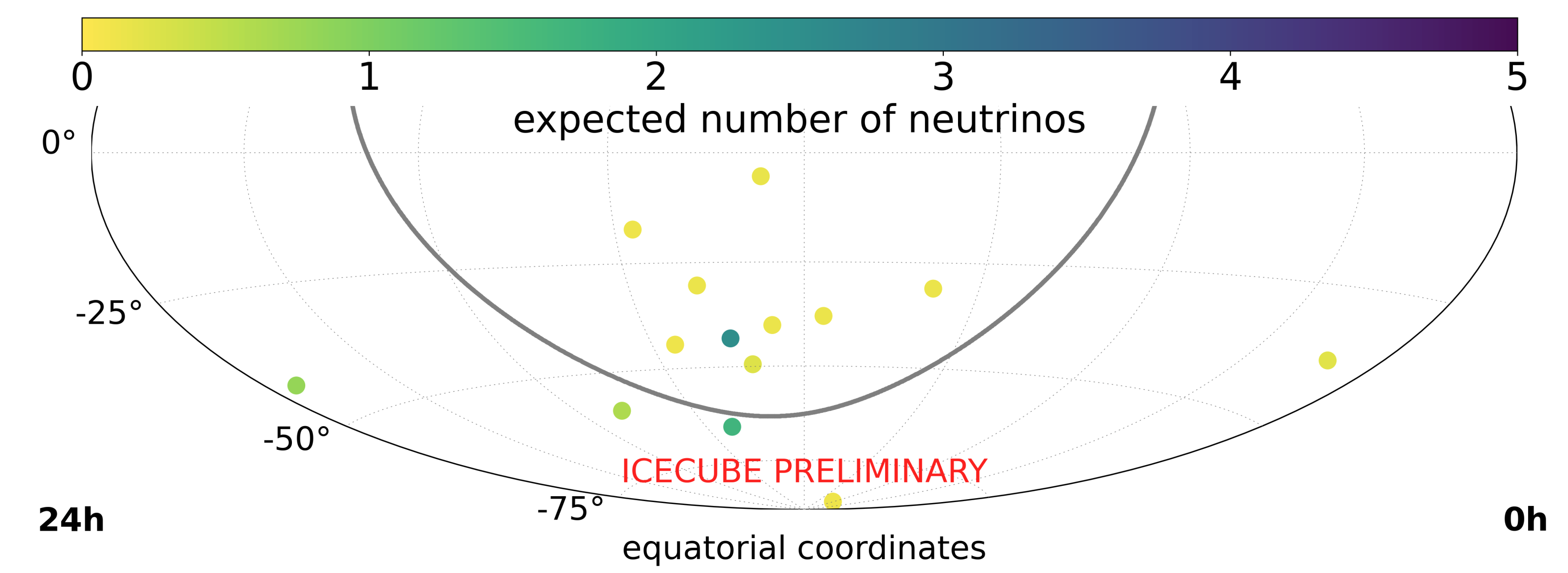}
\caption{\textbf{Distribution of the Seyfert Galaxies} in the southern Sky, that are selected for this work. The color code represents the expected number of neutrinos using the disk-corona flux model. The gray line denotes the location of the galactic plane. The skymap uses equatorial coordinates.}
\label{fig:skymap}
\end{figure}

\noindent BASS~\cite{Ricci:2017dhj} is an all-sky study of Swift/BAT detected X-ray AGN. We use the source classifications from the 105-month $Swift$-BAT survey ~\cite{Oh:2013wzc, Oh:2018wzc} to identify the Seyfert galaxies among the Southern sources in BASS. Combining the disc-corona neutrino flux model  \cite{Murase:2019vdl, Kheirandish:2021wkm} with the intrinsic X-ray luminosities ($2-10\,\mathrm{keV}$, absorption corrected), and distances reported by BASS, we compute the expected number of neutrinos in our dataset, accounting for the corresponding IceCube neutrino effective area. Fig. \ref{fig:selection} (left) shows the resulting expected number of neutrinos as a function of the distance of each Seyfert galaxy. We then choose the top 14 Southern Seyfert galaxies for our analysis using the neutrino expectation as a metric.
The neutrino fluxes predicted by the disc-corona model for our selection of Seyfert galaxies is shown Fig. \ref{fig:selection} (right). The two objects with the largest number of expected neutrinos are Centaurus A (Cen A) and Circinus Galaxy, with $2.5$ and $1.7$ neutrinos predicted,  respectively, if all of our assumptions are exactly met. In total, we expect $7.2$ neutrinos from our selection of Seyfert galaxies. The distribution of these sources in the Southern Sky and their expected number of neutrinos are shown in Fig. \ref{fig:skymap}. 

\noindent Cen A differs from the typical Seyfert galaxy in our sample in that it is known for its radio-bright, relativistic jet. While typical, "radio-quiet" Seyfert Galaxies are usually not detected in the gamma-ray bands \cite{Ackermann_2012}, Cen A is known to emit gamma-rays up to the VHE regime \cite{HESS:2018cxr}, which may originate from this jet. Similarly, the origin of the X-ray emission of Cen A is under debate, with a jet origin being one possibility \cite{Kheirandish:2021wkm}. In our modeling, we have assumed that the X-rays are coronal in nature, thus making the predicted number of neutrinos highly uncertain for this source. As the X-ray emission from Circinus Galaxy is believed to be coronal in origin, the model prediction is more robust, thus making it the most promising source in this analysis. Due to its proximity to the Milky Way, Circinus Galaxy is, in principle, visible as an extended object. Nevertheless, neutrino emission from the coronal region of its AGN would appear as a point-like source to IceCube. 



\section{Analysis Methodology}\label{sec_method}
\noindent Following the analysis methods of the companion analysis in the Northern Sky \cite{IceCube:2023northseyfert}, we will perform two types of searches: a catalog and a stacking search.\\ 

\noindent \textbf{Catalog Searches}\hspace{2em}Here, we individually analyze each of the 14 sources in our selection for an excess of point-like neutrino emission over backgrounds from atmospheric neutrinos, the diffuse astrophysical neutrino flux \cite{diffus:nue/nutau, diffuse:numu}, and neutrinos from our own galaxy \cite{Abbasi:2023bvn}. We also perform a binomial test \cite{IceCube:2018ndw}, a technique that allows us to combine, in a statistical manner, potential excess neutrino emission from multiple sources in a relatively model-independent way, i.e., without having to make any prior assumptions about the relative strength of the emission across the sources in our catalog. Here, we are primarily interested in results assuming the disc-corona flux model, but we will also perform the analysis assuming the canonical power-law flux to guard against model miss-specifications and enable easier comparisons to other IceCube works.\\

\noindent \textbf{Stacking Search}\hspace{2em}Assuming all of our model assumptions, including the intrinsic X-ray fluxes estimated in BASS, are correct, a stacking search, the joint analysis of all sources in which sources are weighted according to their expected number neutrinos, provides the most statistical power in identifying coronal neutrino emission from our set of sources. However, the increase in statistical power over the binomial search comes at the cost of increased impact of systematic errors in our model assumptions (or inputs). Because the strong X-ray emission of Cen A may be related to its jet activity rather than coronal processes, we exclude this source from our stacking search. Otherwise, this source would make the dominant contribution to this type of analysis, thereby propagating this systematic uncertainty to the other sources in the stack. The stacking search is performed only for the disc-corona model, i.e., the power-law flux is not tested in this part of our work.\\

\noindent All searches rely on the standard unbinned maximum likelihood formalism \cite{Braun:2008bg, IceCube:2022der}, which considers the reconstructed event directions and energies, as well as estimated per-event angular uncertainties as the basis of the statistical method. While the generic power-law flux introduces two fit parameters, the mean number of signal events $n_s$ and spectral index $\gamma$, our disc-corona flux model (including all assumptions) fully specifies the expected neutrino observations, and the model-based search could therefore be performed without any free parameters. We acknowledge the existence of systematic uncertainties in the estimated X-ray fluxes, especially for Compton-thick AGN, the estimated source distances as well as other parameters intrinsic to the model. Changing, for example, the assumed cosmic-ray to thermal pressure ratio \cite{Kheirandish:2021wkm}, would change the normalization of the predicted neutrino flux. For each source, we therefore treat the normalization, corresponding to $n_s$, as one free parameter. In the stacking search, the total normalization is kept free, while the relative contributions of the different sources are kept fixed at the nominal predictions.\\
\begin{figure}[tbp!]
\centering
\includegraphics[width=0.6\textwidth]{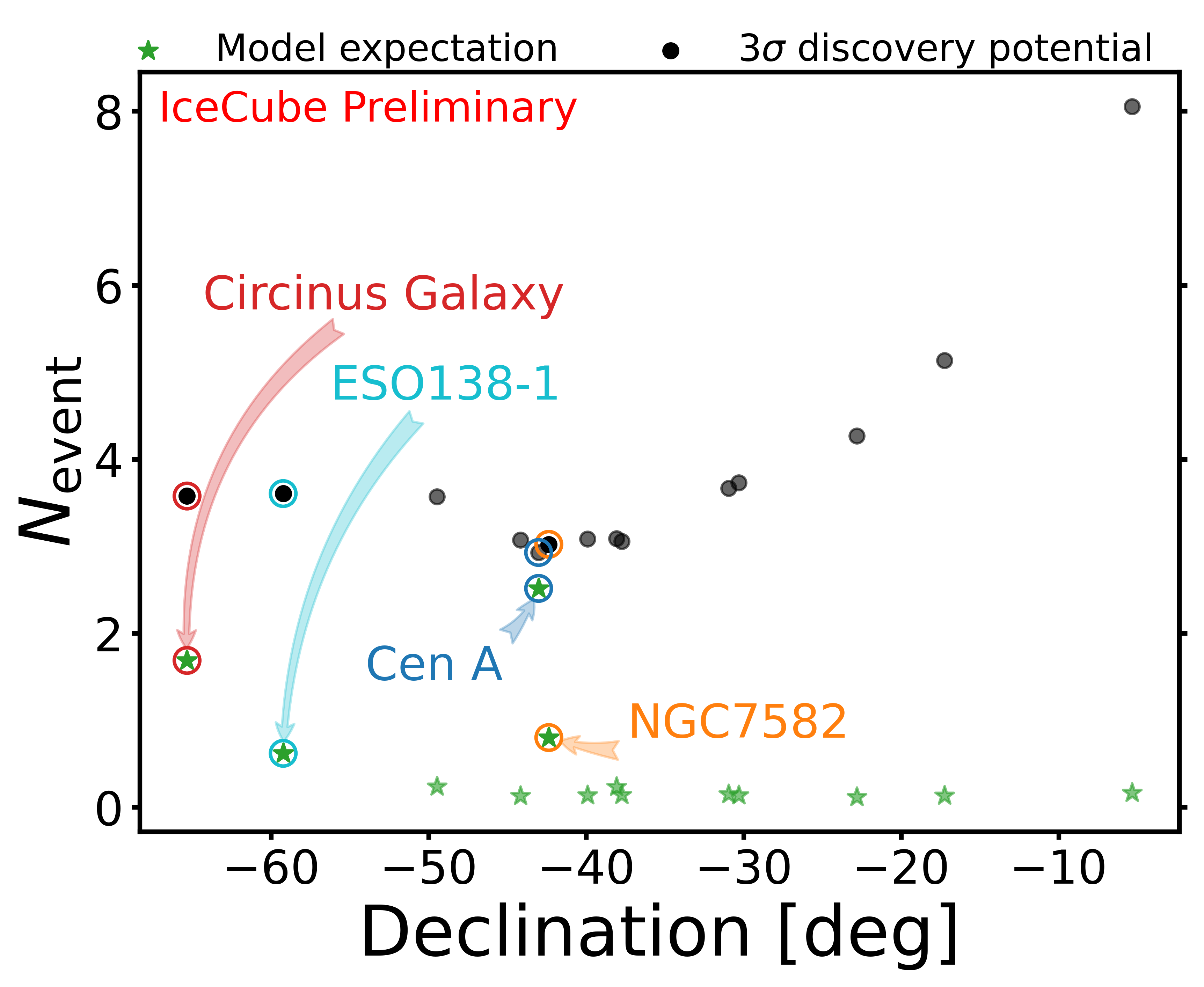}
\caption{Expected numbers of events from the model (green stars) are compared with the 3$\sigma$ discovery potentials (black dots). The four brightest sources, Cen A, Circinus Galaxy, NGC 7582, and ESO 138-1 are highlighted.}
\label{catalog}
\end{figure}

\noindent Because we are analyzing the Southern Sky, we need to account for four types of backgrounds: atmospheric muons, atmospheric neutrinos, astrophysical neutrinos, and neutrinos from the Milky Way. The recent NGC 1068 result deployed MC simulations to model the backgrounds, because it naturally avoids signal contamination in the background data, while at the same time being able to fully sample the observable space \cite{IceCube:2022der}. Here, in the Southern Sky, systematic uncertainties in the modeling of the muon background as well as its correlation with atmospheric neutrinos, the so-called self-veto effect \cite{Arguelles:2018awr}, prevent us from using the same method. Instead, we rely on the well-tested data scrambling technique. Experimental data is randomized in the right ascension coordinate, in which the detector response, as well as atmospheric and isotropic astrophysical backgrounds, appear uniform. However, this is not true for Galactic neutrinos. In addition, the technique suffers from signal contamination, as potential point-like neutrino emission would be added to the background model, thereby reducing the sensitivity of the search. To alleviate both issues, we modify the standard scrambling method as follows: we first apply a masking technique. Events that fall into the vicinity of the brightest Seyfert galaxies (Cen A, Circinus, NGC 7582, and ESO 138-1, circle of radius 7$^{\circ}$), or the Galactic plane (Galactic latitude $\pm10^{\circ}$ ), are removed from the dataset. Scrambling the remaining data, while accounting for the missing solid angle, then gives an excellent approximation to the atmospheric and diffuse astrophysical backgrounds. Finally, we add simulated events from the Galactic plane into this hybrid dataset, using the best-fit normalization reported in \cite{Abbasi:2023bvn} for the neutrino flux expected from the diffuse galactic gamma-ray flux ($\pi^0$ template) measured by Fermi-LAT ~\cite{paper:pi0}. Following \cite{Abbasi:2023bvn}, we assume the corresponding neutrino flux to follow a power-law with a spectral index of 2.7. Using this hybrid dataset, we obtain the background pdfs required by the likelihood formalism. It also allows us to generate arbitrarily many background-only datasets but with strongly reduced signal contamination. The latter is necessary to compute background test-statistics distributions, a crucial ingredient for the likelihood ratio test. The original, unscrambled, unmasked experimental data is used to evaluate the likelihood function on the experimental data during the data fitting process.

\begin{figure}[tbp!]
\centering
\includegraphics[width=0.6\textwidth]{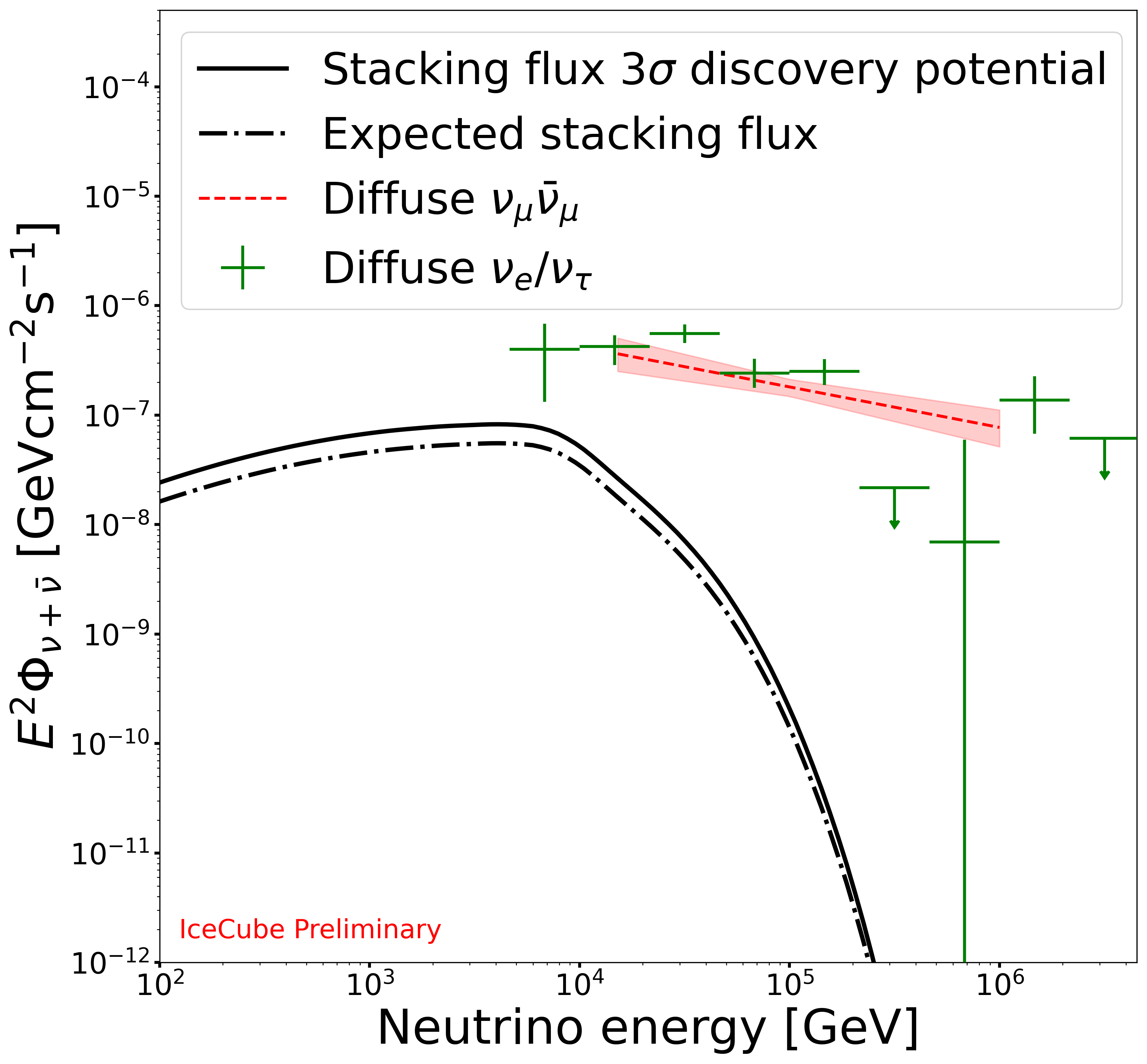}
\caption{The 3$\sigma$ stacking discovery potential of the selected sources from the disk-corona model (black solid), to be compared with the expected stacked flux (black dash-dotted). The measured diffuse astrophysical fluxes of $\nu_\mu$ and $\nu_e/\nu_\tau$ are also shown.   }
\label{fig:stacking_result}
\end{figure}

\section{Projected Analysis Performance}\label{sec_result}
\noindent We measure the performance of the analysis using the sensitivity and discovery potential as metrics. The sensitivity is defined as the median value of the upper limits that the experiment could obtain in a background-only universe. The discovery potential corresponds to the flux needed to make a $5\sigma$ discovery with $50\,\%$ probability. Because no analytic closed-form expression for the disc-corona flux model exists, we will quote these metrics in units of the number of expected events. A comparison between the sensitivity and discovery potential for individual objects in our catalog search and the expected number of events according to the disc-corona model is shown in Fig. \ref{catalog}. According to nominal model assumptions, identification of individual sources are not very likely. However, the collective neutrino emission could provide a strong signal for a stacking analysis. The sensitivity and discovery potential of the stacking search are shown in Fig. \ref{fig:stacking_result}. If all our assumptions (e.g., the intrinsic X-ray fluxes by BASS, the disc corona model with intrinsic parameters tuned to NGC 1068) are exactly met, the $3\sigma$ discovery potential corresponds to $7.0$ neutrinos (Cen A excluded), i.e., $\sim 150\,\%$  of the expected neutrino flux. 

\section{Summary and Outlook}\label{sec3}
\noindent Motivated by the recently reported evidence for neutrino emission from the Seyfert galaxy NGC 1068 in the Northern Sky, and the fact that the majority of the bright Seyfert galaxies reside in the Southern hemisphere,  we have developed a search for neutrino emission from the intrinsically brightest-in-X-ray ($2-10\,\mathrm{keV}$) Seyfert galaxies in the Southern Sky. This analysis offers new opportunities to investigate neutrino emission from some of the primary candidates for IceCube's high-energy neutrino flux. The analysis uses starting track events in IceCube, while the selection of sources rests on the information provided by BASS. According to the disc-corona model for neutrino emission from AGN, the Circinus Galaxy is the most promising target in our search because of its high intrinsic X-ray flux that is believed to originate from the corona of the AGN. Because of the large atmospheric muon background in the Southern Sky, IceCube's performance in is reduced compared to the North, where the Earth shields the detector. Thus, a robust identification of individual sources in our catalog with IceCube is challenging because the number of neutrinos expected from each source is small. Combining possible signals using a likelihood-based stacking method is more promising. \\ 
A closely related analysis, using essentially the same methods, has recently been applied to 27 X-ray bright Seyfert galaxies in the Northern Sky (excluding $\mathrm{NGC}\,1068$) \cite{IceCube:2023northseyfert}. While the data for $\mathrm{NGC}\,4151$ and $\mathrm{CGCG}\,420-015$ appears inconsistent with background expectations at $2.7\,\sigma$ significance, the stacking search constrains the model flux at the $\sim30\,\%$ level of the model prediction in the optimistic scenario. The extent to which the properties of NGC 1068 within the disc-corona model apply to the population of Seyfert galaxies therefore remains very much an open topic \cite{IceCube:2023northseyfert}. The work presented here focuses on a distinct set of sources and is thus bound to provide important additional information about potential neutrino production mechanisms in AGN. By extending the analysis presented here to also include IceCube cascade data \cite{Abbasi:2023bvn} in the Southern Sky, we intend to double the sensitivity of this search in future work.

\bibliographystyle{ICRC}
\bibliography{references}

%

\clearpage

\section*{Full Author List: IceCube Collaboration}

\scriptsize
\noindent
R. Abbasi$^{17}$,
M. Ackermann$^{63}$,
J. Adams$^{18}$,
S. K. Agarwalla$^{40,\: 64}$,
J. A. Aguilar$^{12}$,
M. Ahlers$^{22}$,
J.M. Alameddine$^{23}$,
N. M. Amin$^{44}$,
K. Andeen$^{42}$,
G. Anton$^{26}$
% ...
\\
\\
% ...
$^{12}$ Universit{\'e} Libre de Bruxelles, Science Faculty CP230, B-1050 Brussels, Belgium \\
% ...
$^{17}$ Department of Physics, Loyola University Chicago, Chicago, IL 60660, USA \\
$^{18}$ Dept. of Physics and Astronomy, University of Canterbury, Private Bag 4800, Christchurch, New Zealand \\
% ...
$^{22}$ Niels Bohr Institute, University of Copenhagen, DK-2100 Copenhagen, Denmark \\
$^{23}$ Dept. of Physics, TU Dortmund University, D-44221 Dortmund, Germany \\
% ...
$^{26}$ Erlangen Centre for Astroparticle Physics, Friedrich-Alexander-Universit{\"a}t Erlangen-N{\"u}rnberg, D-91058 Erlangen, Germany \\
% ...
$^{40}$ Dept. of Physics and Wisconsin IceCube Particle Astrophysics Center, University of Wisconsin{\textendash}Madison, Madison, WI 53706, USA \\
% ...
$^{42}$ Department of Physics, Marquette University, Milwaukee, WI, 53201, USA \\
% ...
$^{44}$ Bartol Research Institute and Dept. of Physics and Astronomy, University of Delaware, Newark, DE 19716, USA \\
% ...
$^{63}$ Deutsches Elektronen-Synchrotron DESY, Platanenallee 6, 15738 Zeuthen, Germany  \\
$^{64}$ Institute of Physics, Sachivalaya Marg, Sainik School Post, Bhubaneswar 751005, India \\
% ...

\subsection*{Acknowledgements}

\noindent
The authors gratefully acknowledge the support from the following agencies and institutions:
USA {\textendash} U.S. National Science Foundation-Office of Polar Programs,
U.S. National Science Foundation-Physics Division,
U.S. National Science Foundation-EPSCoR,
Wisconsin Alumni Research Foundation,
Center for High Throughput Computing (CHTC) at the University of Wisconsin{\textendash}Madison,
Open Science Grid (OSG),
Advanced Cyberinfrastructure Coordination Ecosystem: Services {\&} Support (ACCESS),
Frontera computing project at the Texas Advanced Computing Center,
U.S. Department of Energy-National Energy Research Scientific Computing Center,
Particle astrophysics research computing center at the University of Maryland,
Institute for Cyber-Enabled Research at Michigan State University,
and Astroparticle physics computational facility at Marquette University;
Belgium {\textendash} Funds for Scientific Research (FRS-FNRS and FWO),
FWO Odysseus and Big Science programmes,
and Belgian Federal Science Policy Office (Belspo);
Germany {\textendash} Bundesministerium f{\"u}r Bildung und Forschung (BMBF),
Deutsche Forschungsgemeinschaft (DFG),
Helmholtz Alliance for Astroparticle Physics (HAP),
Initiative and Networking Fund of the Helmholtz Association,
Deutsches Elektronen Synchrotron (DESY),
and High Performance Computing cluster of the RWTH Aachen;
Sweden {\textendash} Swedish Research Council,
Swedish Polar Research Secretariat,
Swedish National Infrastructure for Computing (SNIC),
and Knut and Alice Wallenberg Foundation;
European Union {\textendash} EGI Advanced Computing for research;
Australia {\textendash} Australian Research Council;
Canada {\textendash} Natural Sciences and Engineering Research Council of Canada,
Calcul Qu{\'e}bec, Compute Ontario, Canada Foundation for Innovation, WestGrid, and Compute Canada;
Denmark {\textendash} Villum Fonden, Carlsberg Foundation, and European Commission;
New Zealand {\textendash} Marsden Fund;
Japan {\textendash} Japan Society for Promotion of Science (JSPS)
and Institute for Global Prominent Research (IGPR) of Chiba University;
Korea {\textendash} National Research Foundation of Korea (NRF);
Switzerland {\textendash} Swiss National Science Foundation (SNSF);
United Kingdom {\textendash} Department of Physics, University of Oxford.

\end{document}